\documentclass[a4paper]{article}

\usepackage{INTERSPEECH2021}
\usepackage{url}

\title{Digital Einstein Experience: Fast Text-to-Speech for Conversational AI}

\name{Joanna Rownicka, Kilian Sprenkamp, Antonio Tripiana, Volodymyr Gromoglasov, Timo P Kunz}
\address{Aflorithmic Labs Ltd.}
\email{\{joanna, kilian, antonio, volodymyr, timo\}@aflorithmic.ai}

\begin{document}

\maketitle
\begin{abstract}
We describe our approach to create and deliver a custom voice for a conversational AI use-case. More specifically, we provide a voice for a Digital Einstein character, to enable human-computer interaction 
within the digital conversation experience. To create the voice which fits the context well, we first design a voice character and we produce the recordings which correspond to the desired speech attributes. We then model the voice. Our solution utilizes Fastspeech 2 for log-scaled mel-spectrogram prediction from phonemes and Parallel WaveGAN to generate the waveforms. The system supports a character input and gives a speech waveform at the output.  We use a custom dictionary for selected words to ensure their proper pronunciation. Our proposed cloud architecture enables for fast voice delivery, making it possible to talk to the digital version of Albert Einstein in real-time. 
\end{abstract}

\section{Introduction}



Spoken dialog systems find application in everyday digital assistants, as well as in conversational social commerce. Some of the use-cases include customer service, marketing, support, coaching, entertainment and education. Text-to-Speech (TTS) is a central component of conversational systems, as it enables human-computer interaction. With speech as an interface between humans and machines, the communication becomes natural, contributing to an enhanced user experience. Customization of the voice is an important aspect of designing a voice suitable for the use-case. The voice characteristics (pitch, tone, pace, rhythm, resonance, texture, inflection, etc.) need to fit the context in which the voice will be used.

In this paper, we describe a TTS system that was developed to provide a custom voice for the Digital Einstein chatbot. The voice was designed such that it meets the expectations of an interacting user. First and foremost, we aimed at creating a good imitation of Albert Einstein's voice. Secondly, we focused on fast voice delivery, to enable dynamic interaction with the user. The proposed system is a demonstration of how creating a custom voice for a spoken dialog system can enrich the overall user experience.

\section{System description}
The Digital Einstein Experience\footnote{\url{https://einstein.digitalhumans.com/}} is an example of a conversational AI system. It consists of video and audio components. The chatbot was created for educational and entertainment purposes. The user can chat to the digital version of Albert Einstein on a variety of topics, e.g. his life, science, technology. It is possible to type or say any question or select one from a list of suggestions. The answers are provided by WolframAlpha\footnote{\url{https://www.wolframalpha.com/}}. Moreover, the Digital Einstein can quiz the user on a selected topic. Quiz data is provided by OpenTriviaDB\footnote{\url{https://opentdb.com/}}. This paper focuses on the audio, i.e. the voice component of the Digital Einstein Experience.




We first describe our approach to create the TTS model. Then, we depict the cloud architecture used for Synchronous TTS delivery, which allowed for subsecond voice creation.

\begin{figure}[t]
    \centering
    \includegraphics[scale=0.135]{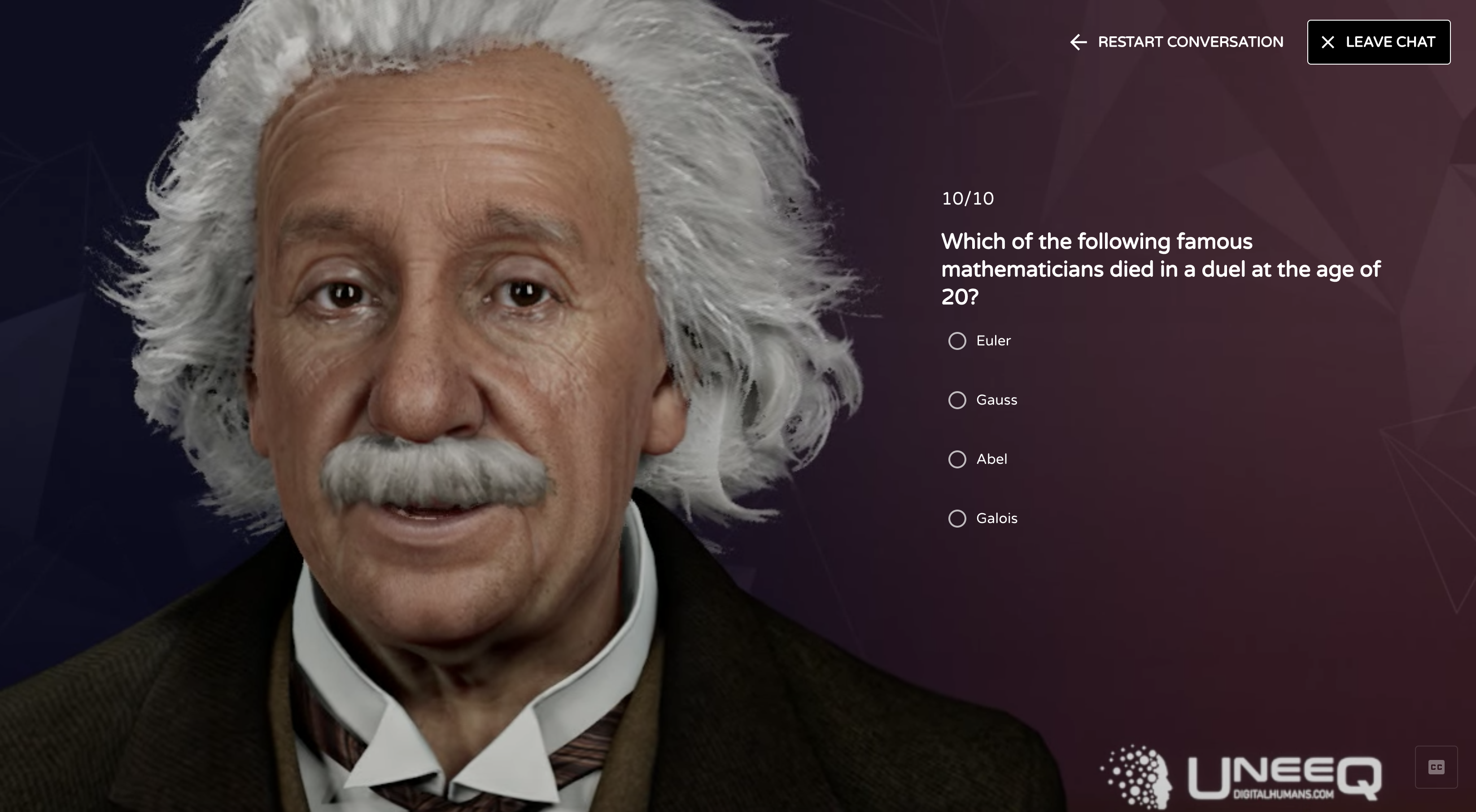}
    \caption{Interface for the Digital Einstein Experience.}
    \label{fig:interface}
\end{figure}

\subsection{Voice creation}

We started by defining the attributes of the voice that we would like to create. The main requirements for a recreation of Einstein's voice were: German accent, rather high pitch, slow pace. We refer the reader to our blog post to learn more about designing the digital Einstein character\footnote{\url{https://www.aflorithmic.ai/post/creating-einsteins-voice}}.

\begin{figure*}
    \centering
    \includegraphics[scale=0.6]{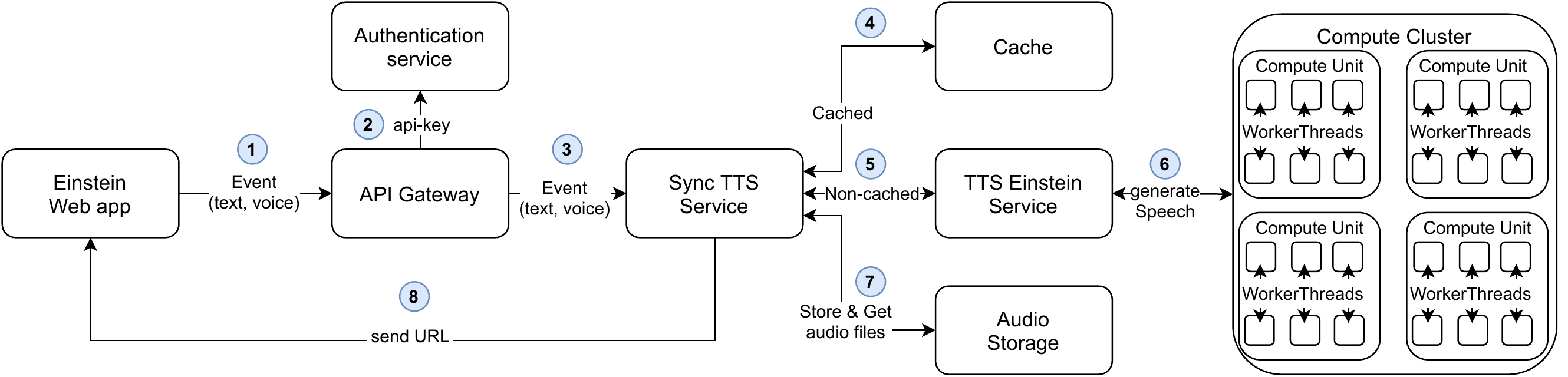}
    \caption{Cloud architecture for Synchronous TTS.}
    \label{fig:architecture}
\end{figure*}

The recordings were provided by a professional voice actor. We applied Google's WebRTC Voice Activity Detection (VAD) algorithm to exclude silent frames. Utterances and text were aligned manually. We used the utterances of length $0.1-40$ seconds and we extracted the $80$-dimensional FBANK acoustic features (i.e. log-scaled mel-spectrograms) with a $2048$-point FFT using a Hanning window of width $1200$, with a hop of $300$. The frequency range for the feature extraction was $80-7600$Hz.

The text was normalized, transforming grapheme sequences into phoneme sequences. We used the CMU dictionary and a neural G2P model to do the grapheme-to-phoneme conversion. We also implemented a custom lexicon for Einstein's voice where pronunciations for selected words can be inserted manually. In this dictionary some German words and phrases were included; e.g. it enabled Einstein to greet the users in German. Custom pronunciations are prioritized over the ones generated with a G2P module.

We used FastSpeech 2 architecture~\cite{fastspeech2} to predict acoustic representations. Before training, we extracted the durations from the target speech waveform with the use of a Tacotron 2 model~\cite{tacotron2}. We also extracted pitch and energy which are needed as conditional inputs for FastSpeech 2 model training. We used token-averaged pitch and energy, similarly as in the FastPitch implementation~\cite{fastpitch}, and a range of $80-400$Hz for pitch extraction. 
Adding variance information such as pitch and energy as input improves the naturalness of the model.

We used Parallel WaveGAN~\cite{pwg} to generate speech waveforms from predicted acoustic features at inference time. This distillation-free and non-autoregressive approach allowed for a fast speech generation without performance degradation, compared to the best distillation-based frameworks~\cite{parallelwavenet}.

\subsection{Cloud architecture}

The cloud architecture used for Synchronous (Sync) TTS is presented in Fig.~\ref{fig:architecture}. The Einstein web app (external system) is an independent system which processes the questions and answers from/to the users in the form of text. In order to generate speech, the Einstein web app sends an HTTPS request - containing the text and the desired voice - to the API Gateway (step~1). The API Gateway is a microservice that ensures the request is coming from a trusted source by checking the request's api-key against an authentication service (step 2). It also redirects the request based on the destination URL path (step 3) to the Sync TTS service.

The Sync TTS service holds the logic to convert Text-to-Speech. It receives the request event with text and the voice selected at the input, and returns an audio file at the output. First, the Sync TTS service checks if the event was produced before, in order to speed up the retrieval process. Two different scenarios can be derived from this check:
\begin{enumerate}
    \item The audio file is cached (step 4): In the case that the file was already produced, the cache will return the URL of the audio file in the audio storage to the Einstein web app. In this scenario, a call to the TTS Einstein Model is not needed which speeds up the TTS retrieval process.
    \item The audio file is not cached (steps 5 and 6): if the file is not cached, the process will continue to the TTS Einstein Service to get a response.
\end{enumerate}

In the latter case, the Sync TTS service will ping the TTS Einstein Service (step 5). In this step, the model produces a speech file from the text provided, and returns it to the Sync TTS service.

In order to serve the Einstein model, we are using an open-source model server. It allows to parallelize API requests across multiple threads running on a single compute instance. Each of the threads keeps the model warm-started, further increasing the inference speed by eliminating the need to load the model with every call. By running multiple instances of the model server on our auto-scaling compute cluster (step 6), it is possible to process multiple speech synthesis tasks in parallel. 

Once the speech file is produced, the Sync TTS service stores the audio in a cloud object storage (step 7), and generates a URL to be returned to the user. Before returning the URL to the user, the Sync TTS service writes a new item to the cache, storing the URL, and the exact combination of text and voice. The user then receives the URL and renders speech in the application (step 8).

\section{Conclusions}


We presented our approach to create a purpose built voice and serve it with low latency for an interactive digital experience. To create the voice for the Albert Einstein character, we used purpose-made recordings. We adopted state-of-the-art TTS techniques to generate a high quality voice. We also created the infrastructure to accommodate subsecond delivery of voice responses.

\section{Acknowledgements}

We would like to thank everyone at Aflorithmic Labs for making this project possible. We are also grateful to UneeQ for giving us the opportunity to complement one of their digital humans.

\bibliographystyle{IEEEtran}

\bibliography{mybib}

\end{document}